\documentclass[pra,twocolumn,superscriptaddress]{revtex4-1}
\usepackage{makeidx}
\usepackage[utf8]{inputenc}
\usepackage{graphicx}
\usepackage{amsmath}
\usepackage{mathrsfs}
\usepackage{graphicx}
\usepackage{subfigure}
\usepackage{dcolumn}
\usepackage{bm}
\usepackage{bbm}
\usepackage{color}
\usepackage{esint}
\usepackage{lineno}
\usepackage[breaklinks,
            colorlinks,
            urlcolor=blue,
            linkcolor=blue,
            anchorcolor=blue,
            citecolor=blue]{hyperref}

\renewcommand{\section}[1]{{\par\it #1.---}\ignorespaces}

\begin{document}

\title{Optical signatures of Mott-superfluid transition in nitrogen-vacancy centers coupled to photonic crystal cavities}
\author{Jia-Bin You}
\affiliation{State Key Laboratory of Magnetic Resonance and Atomic and Molecular Physics, Wuhan Institute of Physics and Mathematics, Chinese
Academy of Sciences, Wuhan 430071, China}
\affiliation{Department of Electronics and Photonics, Institute of High Performance Computing, 1 Fusionopolis Way, 16-16 Connexis,
Singapore 138632, Singapore}
\author{W. L. Yang}
\email{ywl@wipm.ac.cn}
\affiliation{State Key Laboratory of Magnetic Resonance and Atomic and Molecular Physics, Wuhan Institute of Physics and Mathematics, Chinese
Academy of Sciences, Wuhan 430071, China}
\author{G. Chen}
\affiliation{State Key Laboratory of Quantum Optics and Quantum Optics Devices, Institute of Laser Spectroscopy, Shanxi University, Taiyuan 030006,
China}
\author{Z. Y. Xu}
\affiliation{College of Physics, Optoelectronics and Energy, Soochow University, Suzhou 215006, China}
\author{Lin Wu}
\affiliation{Department of Electronics and Photonics, Institute of High Performance Computing, 1 Fusionopolis Way, 16-16 Connexis,
Singapore 138632, Singapore}
\author{Ching Eng Png}
\affiliation{Department of Electronics and Photonics, Institute of High Performance Computing, 1 Fusionopolis Way, 16-16 Connexis,
Singapore 138632, Singapore}
\author{M. Feng}
\email{mangfeng@wipm.ac.cn}
\affiliation{State Key Laboratory of Magnetic Resonance and Atomic and Molecular Physics, Wuhan Institute of Physics and Mathematics, Chinese
Academy of Sciences, Wuhan 430071, China}

\begin{abstract}
We study the phenomenon of controllable localization-delocalization
transition in a quantum many-body system composed of nitrogen-vacancy
centers coupled to photonic crystal cavities, through tuning the different
detunings and the relative amplitudes of two optical fields that drive two
nondegenerate transitions of the $\Lambda $-type configuration. We not only
characterize how dissipation affects the phase boundary using the mean-field
quantum master equation, but also provide the possibility of observing this
photonic quantum phase transition (QPT) by employing several experimentally
observable quantities, such as mean intracavity photon number, density
correlation function and emitted spectrum, exhibiting distinct optical
signatures in different quantum phases. Such a spin-cavity system opens new
perspectives in quantum simulation of condensed-matter and many-body physics
in a well-controllable way.
\end{abstract}

\maketitle

\textit{Introduction.-} Using quantum hybrid systems to simulate
condensed-matter and many-body physics is an exciting frontier of physics
\cite{QM1.1,QM1.2,QM2.1,QM2.2,QM3,QM4,QM5.1,QM5.2}. Among the promising platforms, the scalable
coupled microcavities (superconducting resonators) array doped with quantum
emitters (superconducting qubit) \cite{Lep,wly,Raf,Houck,C1.1,C1.2,C2,C3,C4,C5,C6}
has received much attention. Especially, the integrated photonic networks
based on cavity-emitter coupled systems, such as nitrogen vacancy centers
(NVC) or cold atom interfaced with photonic crystal cavity (PCC) provide a
powerful platform for studying the strongly-correlated states of light and
nonequilibrium quantum phase transition (QPT) \cite%
{C9,CP1,ppc1,ppc2,CP4,CP42}. Despite this remarkable success, realizing
controllable light-matter interaction between electromagnetic quanta and
discrete levels of quantum system in highly scalable devices is a serious
challenge. Additionally, considerably less attention has been paid to the
detection of nonequilibrium QPT phenomena in these hybrid systems, therfore
new important questions arise related to whether it is possible to more
visually observe and detect the optical signature and critical
characteristic of QPT using experimentally observable quantities.
\begin{figure}[tbp]
\includegraphics[width=0.82\columnwidth]{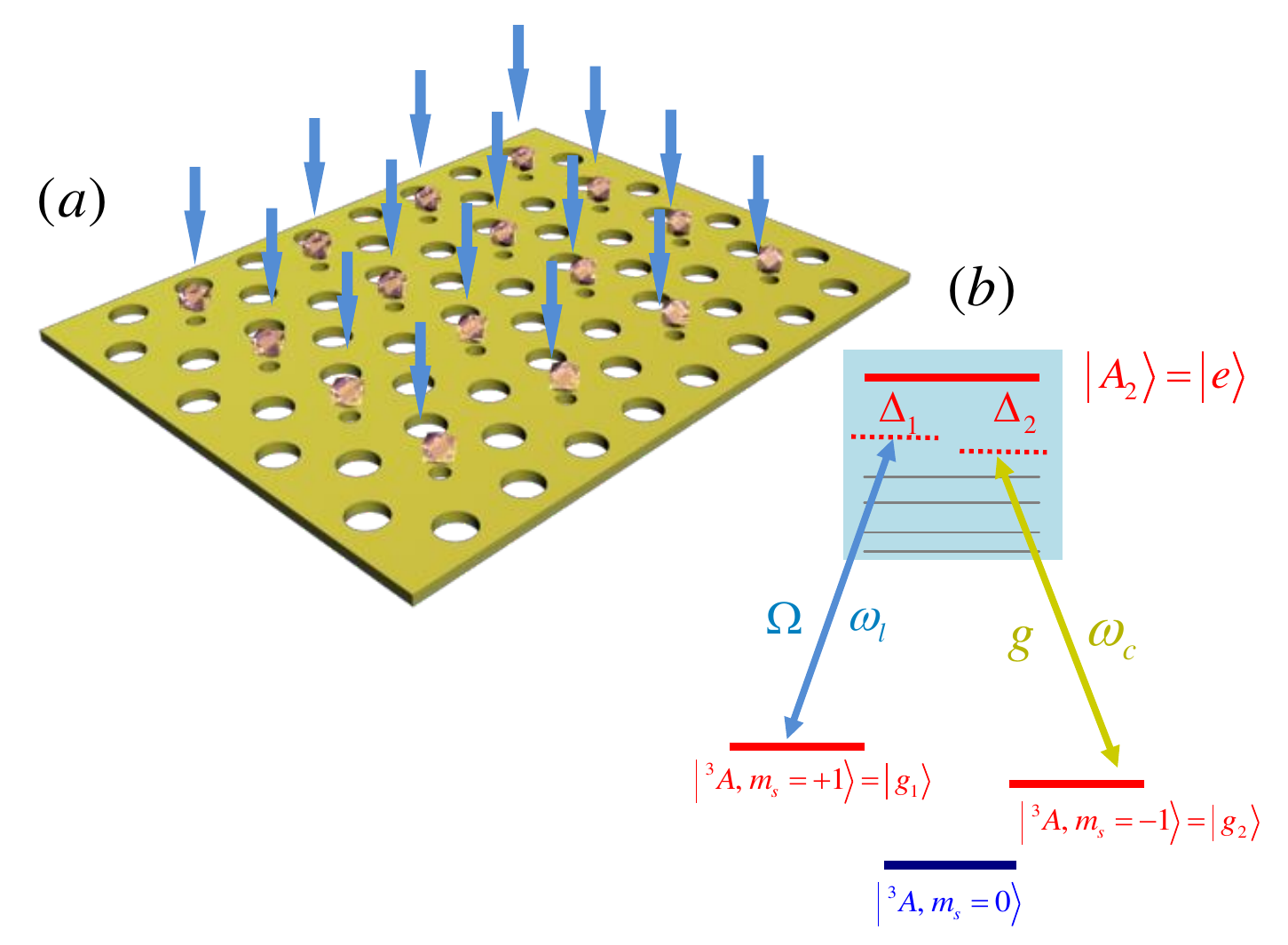}
\caption{(a) The system under consideration consists of a 2D planar PCC
coupled to NVC, where each nanocavity embeds a single NVC, and the parallel
arrows above the NVC denote the classical laser pulses. (b) The level
structure of NVC, where the nanocavity mode and an external control laser
field drive the transitions from the two NVC's ground states $\left\vert
g_{1}\right\rangle $ and $\left\vert g_{2}\right\rangle $ to the excited
state $|e\rangle $.}
\label{3level}
\end{figure}

To this end, the present work focus on a hybrid NVC-PCC system, where each
site composed of a nanocavity and a NVC. Through tuning the polariton states
that are hybridizations of photon and NVC by implementing a dipole-allowed $%
\Lambda $-type transition configuration established by localized tunable
cavity mode and external laser driving, a well-controllable QPT of light
could be realized via full photonic process. In this artificially engineered
hybrid devices,\textit{\ }NVC exhibits excellent optical coherence
properties at ambient conditions and efficient optical control and readout
\cite{NV1,NV2,NV3,NV4,NV5,NV6,NV7,NV8,NV9}. Additionally, a PCC is a
periodic dielectric structure that seeds a localized, tunable cavity mode
around the NVC and controls the propagation of light \cite{CP1}. When many
NVCs are trapped, these dynamically induced cavities mediate coherent
interactions between NVCs \cite{CP2,CP3}. Our proposal is inspired by series
of experimental\emph{\ }demonstrations on strong coupling or quantum
properties of photons in nanophotonic systems with individual solid-state
emitters \cite{ppc1,ppc2,CP4,CP42}, ion \cite{ion} or cold atom \cite%
{CP1,CP5}.

The goal of the present work is twofold. On the one hand, based the
composite NVC-PCC system,\textit{\ }we attempt to demonstrate a controllable
localization-delocalization transition of light by virtue of Raman
transition through adjusting the key parameters for quantum control, which
can be dynamically tuned using available structures of tunable cavity mode
and adjusting external controlling laser fields. Besides exhibiting a
complete picture for the driven-dissipation QPT using the Mean-field quantum
Master equation (MME), the high degree of controllability also\emph{\ }opens
up new possibilities for studying strongly-correlated photon physics in a
well-controlled way. One the other hand, we address the issue of detecting
optical signatures of QPT by calculate the behavior of mean photon number,
photon fluctuation, equal-time correlation function, and emitted spectrum in
complete parameter space. We find that these observable quantities
exhibiting distinct optical signatures in different quantum phases and could
be a good indicator of dissipative QPT.

\textit{The model and Hamiltonian.}- As illustrated in Fig. \ref{3level}(a),
the system under consideration consists of a PCC where each NVC is embedded
in the localized nanocavity with frequency and nonlocal hopping rate tunable
by changing geometrical parameters of the defects \cite{P5.1,P5.2,P5.3,P6.1,P6.2}. The ground
state sublevels are $\left\vert g_{1}\right\rangle =$ $\left\vert
^{3}A,m_{s}=+1\right\rangle $ and $\left\vert g_{2}\right\rangle =\left\vert
^{3}A,m_{s}=-1\right\rangle $ with radiation of $\sigma ^{-}$ and $\sigma
^{+}$ circular polarizations, respectively, whereas the excited level is $%
|e\rangle =\left\vert A_{2}\right\rangle \ $within the spin-orbit excited
state manifold \cite{EEx1,EEx2}. The $\left\vert g_{1}\right\rangle $ spin
state of NVC is linked to $\left\vert A_{2}\right\rangle $ by an
off-resonant laser pulse with detunging $\Delta _{1}$, strength $\Omega $
and frequency $\omega _{l}$, while the transition $\left\vert
g_{2}\right\rangle \longleftrightarrow \left\vert A_{2}\right\rangle $ is
driven by the nanocavity mode with detunging $\Delta _{2}$, strength $g$ and
frequency $\omega _{c}$ \cite{lambda,opt}. This particularly useful $\Lambda
$-type transition was recently employed for spin-photon entanglement
production \cite{Tog}, high-fidelity transfer \cite{NV8}, and holonomic
quantum gate \cite{NV9} in experiments.

Setting the energy of level $|g_{1}\rangle $ to zero, the effective
Hamiltonian at the $i$-th site can be written under a rotating wave
approximation in units of $\hbar $ = 1 as $H_{i}=\omega _{e}|e\rangle
\langle e|+\omega _{2}|g_{2}\rangle \langle g_{2}|+\omega _{c}a^{\dag
}a+\Omega \lbrack |g_{1}\rangle \langle e|e^{i\omega _{l}t}+|e\rangle
\langle g_{1}|e^{-i\omega _{l}t}]+g[|g_{2}\rangle \langle e|a^{\dag
}+a|e\rangle \langle g_{2}|]$, where $\omega _{e}=\omega _{l}+\Delta _{1}$,
and $\omega _{2}=\omega _{l}-\omega _{c}+\Delta _{1}-\Delta _{2}$ with $%
a^{\dag }(a)$ the creation (annihilation) operator of photon. In the
interaction picture the Hamiltonian $H_{i}$ can be reduced as

\begin{equation}
H_{i}^{0}=\sum\nolimits_{i=1,3}\Delta _{i}\sigma _{i}^{+}\sigma
_{i}^{-}+(\Omega \sigma _{1}^{+}+ga\sigma _{2}^{+}+h.c.),  \label{iron_trap}
\end{equation}%
where $\Delta _{3}=\Delta _{1}-\Delta _{2}$, and the lowering and raising
operators are defined as $\sigma _{1}^{+}=|e\rangle \langle g_{1}|$, $\sigma
_{2}^{+}=|e\rangle \langle g_{2}|$ and $\sigma _{3}^{+}=|g_{2}\rangle
\langle g_{1}|$. Note that the total number of excitations $%
N_{i}=a_{i}^{\dag }a_{i}+\sigma _{i,1}^{+}\sigma _{i,1}^{-}+\sigma
_{i,3}^{+}\sigma _{i,3}^{-}$ is the conserved quantity of the system. Adding
the on-site chemical potential and the adjacent-site photon hopping, the
full Hamiltonian for the $2D$ square lattice is given by

\begin{equation}
H=\sum\nolimits_{i}H_{i}^{0}-\sum\nolimits_{\left\langle i,j\right\rangle
}k_{ij}a_{i}^{\dag }a_{i}-\sum\nolimits_{i}\mu _{i}N_{i}.  \label{TotalH}
\end{equation}%
The second term in Eq. (\ref{TotalH}) describes the the nonlocal hopping of
photons between nearest-neighbor nanocavities with the hooping rate $k_{i,j}$%
\emph{.} The third term in Eq.(\ref{TotalH}) describes the on-site chemical
potential with the value $\mu _{i}$ at the $i$-th site, which is
conceptually different from the chemical potential in electronic system. For
convenience to determine the phase diagram, we used the grand canonical
approach considering a situation in which particle exchange with the
surroundings is permitted \cite{Hart}, and we assume zero disorder with $\mu
_{i}=\mu $ and $k_{ij}=k$ for all sites.

\textit{Dissipative QPT}.- Using the mean-field theory \cite{MF.1,MF.2} we decouple
the hopping term as $a_{i}^{+}a_{j}=$ $\left\langle a_{i}^{+}\right\rangle
a_{j}+a_{i}^{+}\left\langle a_{j}\right\rangle -\left\langle
a_{i}^{+}\right\rangle \left\langle a_{j}\right\rangle $ and make a sum over
single sites, the mean-field Hamiltonian can be written as
\begin{equation}
H^{m}=\sum\nolimits_{i}[H_{i}^{0}-zk(a_{i}\psi _{a}^{\ast }+a_{i}^{+}\psi
_{a})+zk\left\vert \psi _{a}\right\vert ^{2}-\mu _{i}N_{i}],
\label{mean_field}
\end{equation}%
where the periodic boundary condition is applied, and $z=4$ is the
coordination number of the lattice.

\begin{figure}[tbp]
\begin{tabular}{cc}
\includegraphics[width=8.6cm]{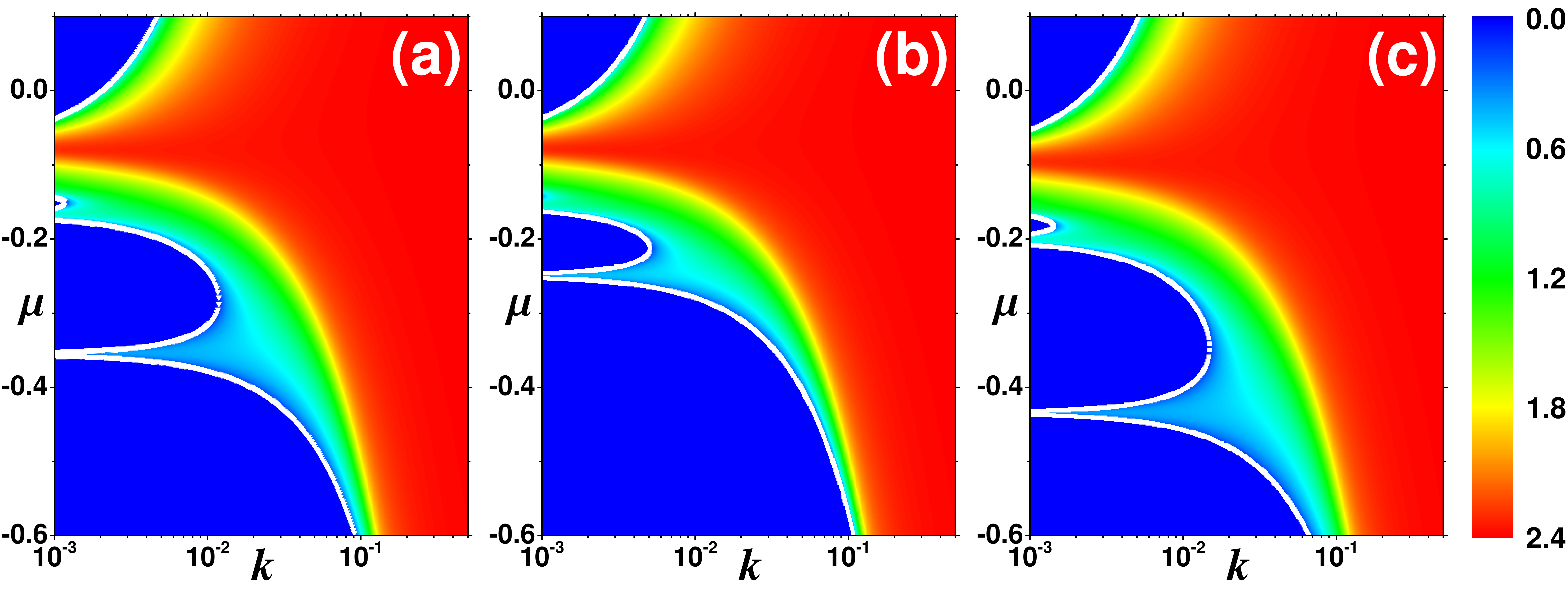} &
\end{tabular}%
\caption{Phase diagrams in the $\protect\mu \sim k$ plane for different
parameters. The common parameters are $g=1$ (used in all the Figures). (a) $\Omega =5$, $\Delta
_{1}=4 $, $\Delta _{2}=-2.5$; (b) $\Omega =5$, $\Delta _{1}=4.3$, $\Delta
_{2}=-2.5$; (c) $\Omega
=4.2$, $\Delta _{1}=3.3$, $\Delta _{2}=-2.1$. The color scale shows the
magnitude of the order parameter $\protect\psi _{a}$.\emph{\ }The white
contour is the SF-MI phase boundary.}
\label{mott-sf_origin}
\end{figure}

We choose the superfluid (SF) phase order parameter $\psi _{a}=\langle
a_{i}\rangle $ (set to be real) to differentiate the different phases.
Minimizing the ground state energy of the Hamiltonian $H^{m}$ (Eq.(\ref%
{mean_field})) with respect to $\psi _{a}$ for different values of $\mu $
and $k$, we obtain the mean field phase diagram/boundary in the $(\mu ,k)$
plane for different tunable parameters, as shown in Fig. \ref{mott-sf_origin}%
. When the on-site large repulsion resulting from laser-assisted spin-cavity
coupling dominates ($k\ll g_{eff}=(\Delta _{1}+\Delta _{2})g\Omega /(2\Delta
_{1}\Delta _{2})$), the system is in the Mott insulator (MI) phase with $%
\psi _{a}=0$, which obeys the $U(1)$ gauge transformation. In the
incompressible MI phase characterized by a fixed number of excitations at
per site with no variance, the photon fluctuations in each nanocavity are
suppressed due to the strong nonlinearity and anharmonicity in the spectrum
originating from the photon blockade effect \cite{Blo}, and the on-site
repulsive interaction between the local photons freezes out hopping and
localizes polaritons at individual lattice sites. By contrast, when the
hopping process with $k\gg g_{eff}$ dominates the dynamics, strong hopping
favours delocalization and condensation of the particles, therefore the
system prefers a $U(1)$-symmetry broken SF phase with $\psi _{a}\neq 0$. In
the compressible SF phase with non-integer polariton number and large
fluctuation, the lowest-energy state of the system is a condensate of
delocalized polariton, and the stable ground state at each site corresponds
to a coherent state of excitations [x]. Therefore, The physical picture
behind is that the MI-SF phase transition results from the interplay between
on-site repulsive interaction and polariton delocalization.

Note that $H^{m}$ is invariant under an $U(1)$ gauge transform: $\psi
_{a}\rightarrow \psi _{a}e^{i\phi }$, $a\rightarrow ae^{i\phi }$, $\sigma
_{1}^{+}\rightarrow \sigma _{1}^{+}e^{-i\phi }$, implying that all odd-order
terms in the expansion of $E_{g}(\psi _{a})$ vanish \cite{PhysRevA.80.023811}%
. Therefore, the ground-state energy has an expression $E_{g}(\psi
_{a})=E_{g}^{0}+(zk+r)|\psi _{a}|^{2}+u|\psi _{a}|^{4}+O(|\psi _{a}|SF^{6})$%
, where $E_{g}^{0}$ is ground state energy of the Hamiltonian $\tilde{H}%
=\sum\nolimits_{i}(H_{i}^{o}-\mu N_{i}).$ From Landau phase transition
theory [x] and second-order perturbation theory, the MI-SF transition occurs
when $zk+r=0$, then we obtain the phase boundary line calculated from $%
r=(zk)^{2}\sum_{m\neq n}|\langle m^{0}|(a+a^{\dag })|n^{0}\rangle
|^{2}/(E_{n}^{0}-E_{m}^{0})$, which is plotted by the white contour in Fig. %
\ref{mott-sf_origin}, where $|n^{0}\rangle $ $(E_{n}^{0})$ is the eigenstate
(eigenenergy) of Hamiltonian $\tilde{H}$.

\begin{figure}[tbp]
\begin{tabular}{cc}
\includegraphics[width=3.5cm]{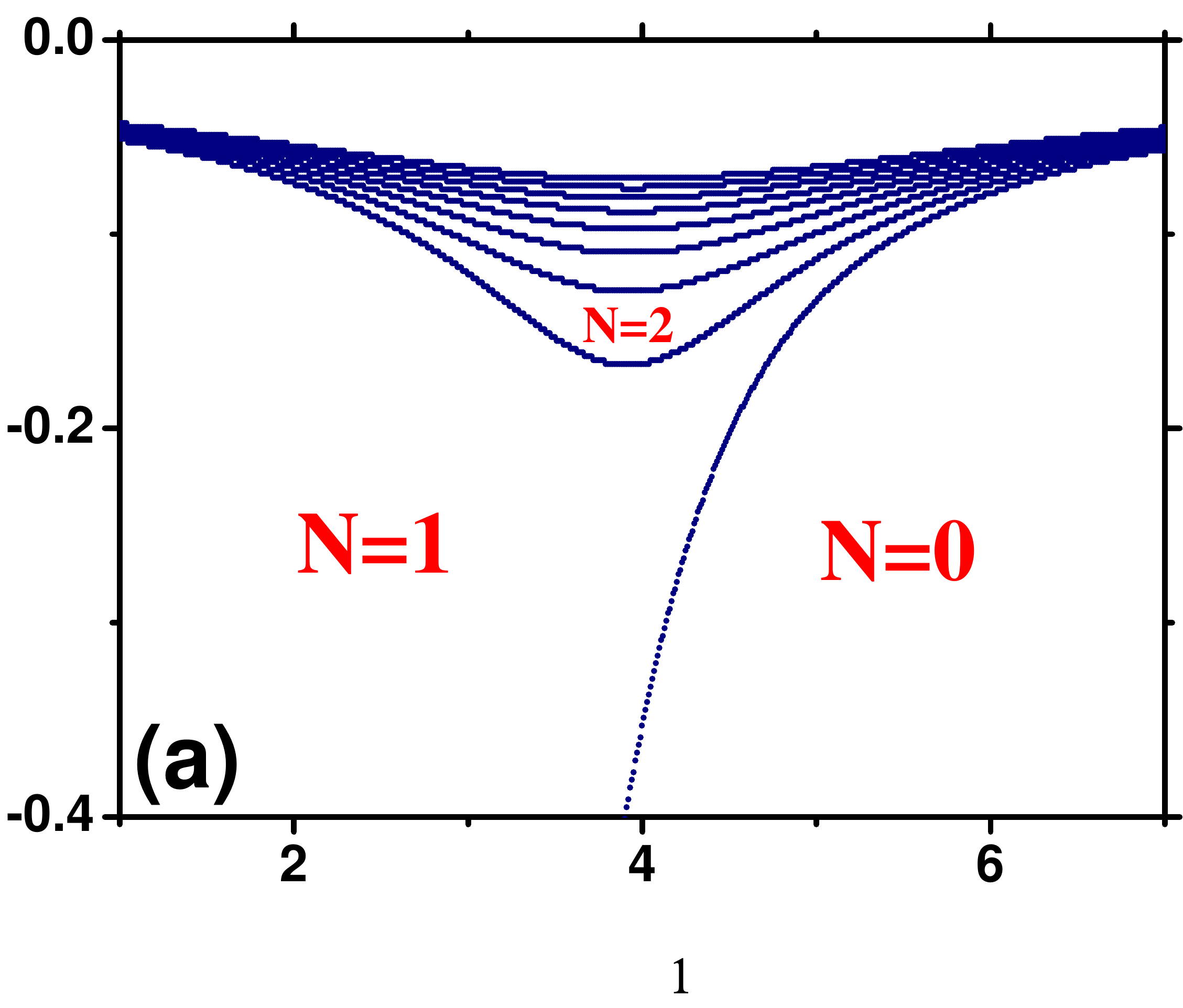} & %
\includegraphics[width=3.5cm]{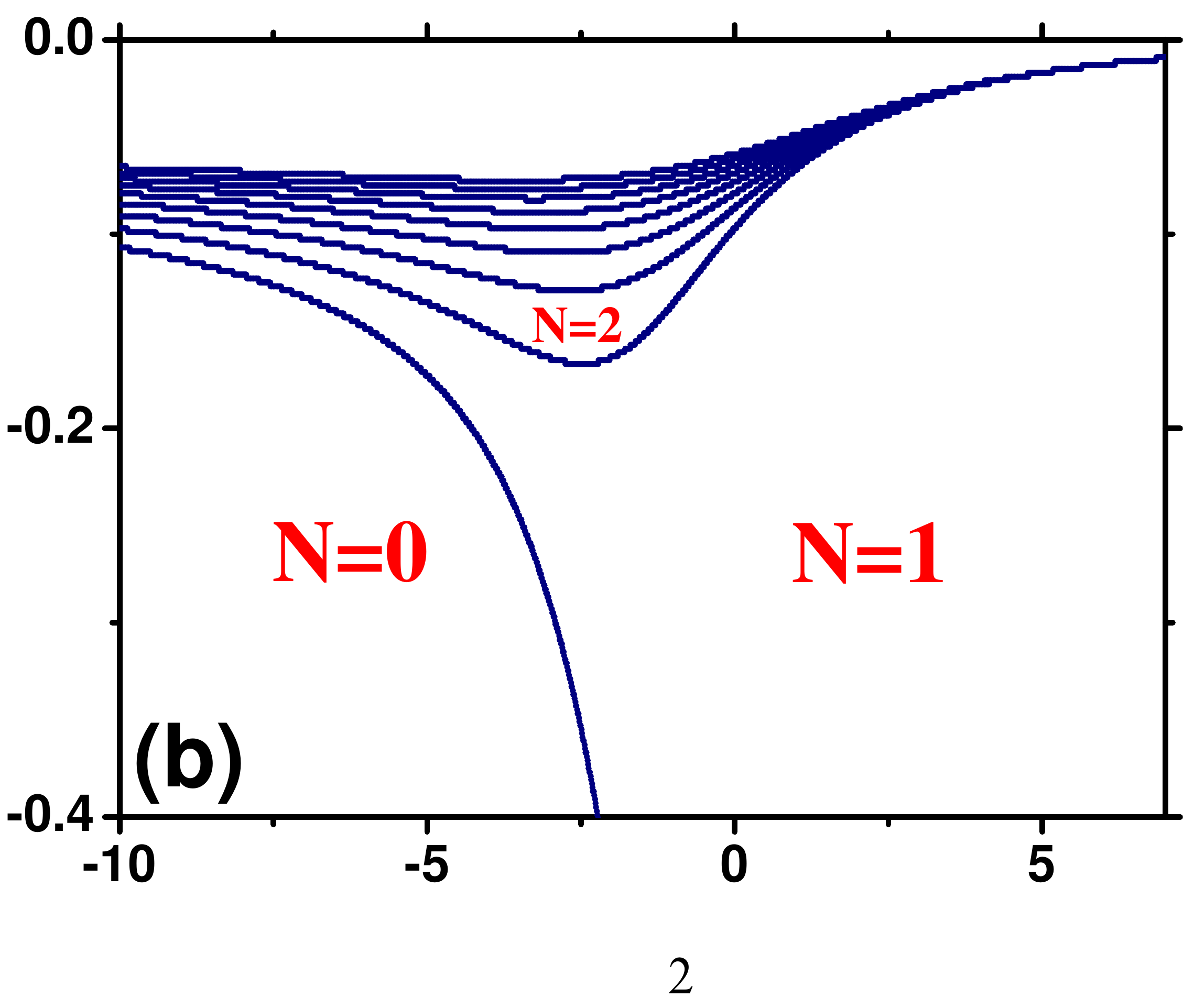} \\
\includegraphics[width=3.5cm]{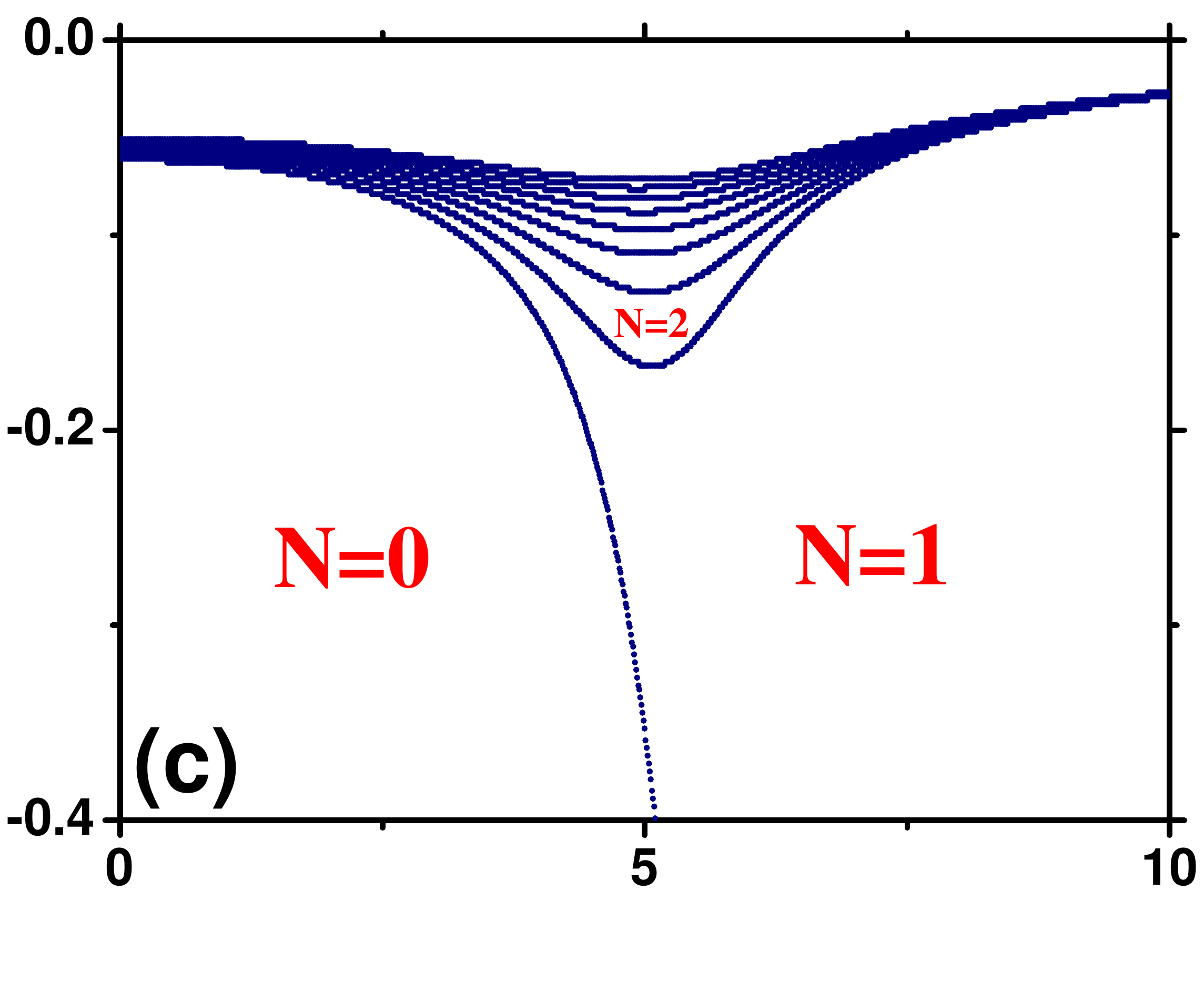} & %
\includegraphics[width=3.5cm]{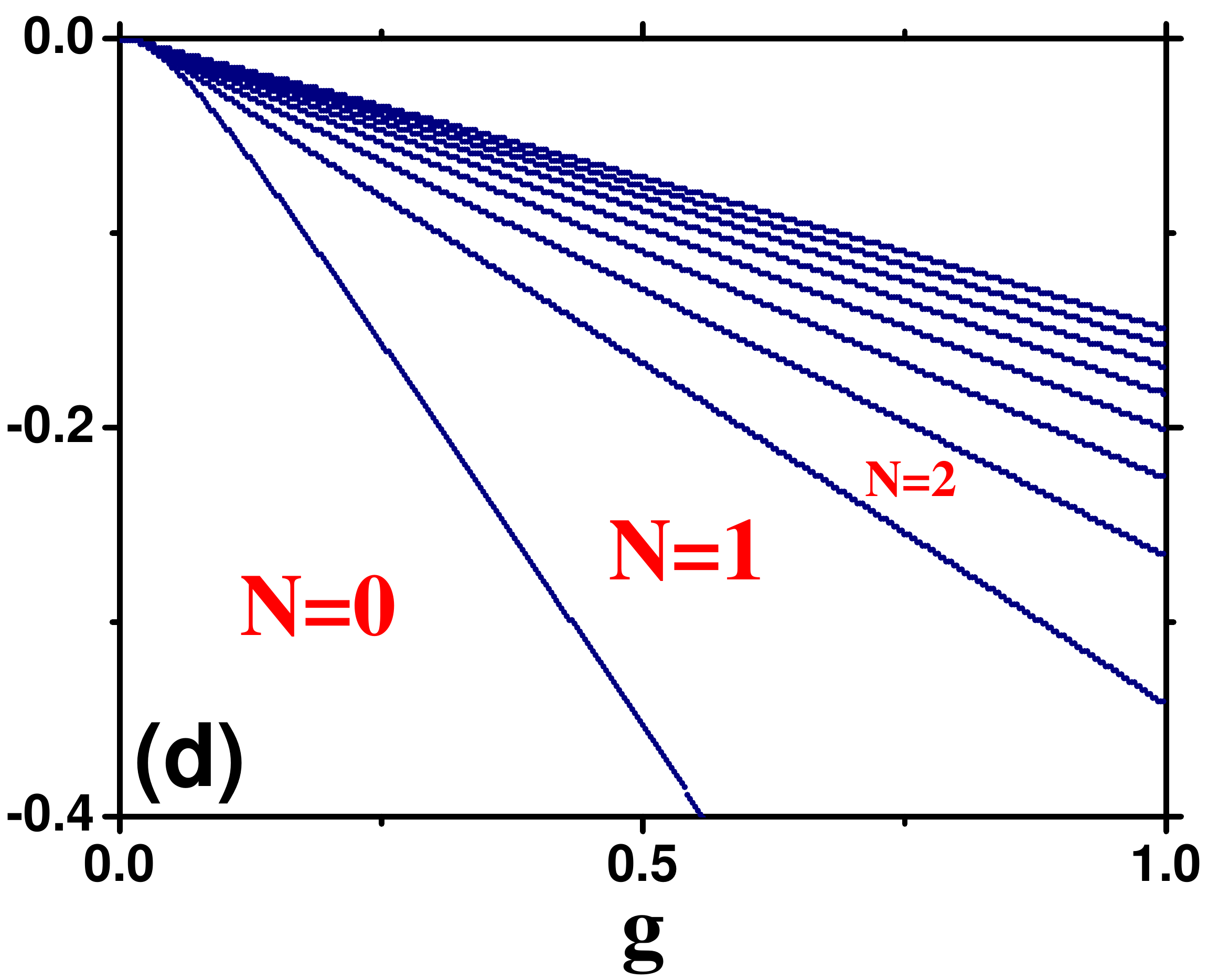} \\
&
\end{tabular}%
\caption{Boundaries between MI lobes in the limit of low tunnelling ($%
k\approx 0$) as a function of detunings and couplings. The parameters in
each figure are the same as Fig. (\protect\ref{mott-sf_origin}a).}
\label{boundary}
\end{figure}

Fig. \ref{mott-sf_origin} tells us\emph{\ }that phase boundary and size of
the MI lobes primarily depend on the ratio of on-site repulsion rate to
hopping rate, and it could also be shifted and changed by adjusting the
controllable parameters \{$\Omega $, $\Delta _{1}$, $\Delta _{2}$\}. We
visualize\emph{\ }the corresponding boundary lines between MI lobes labelled
with different polariton number $\langle {N}\rangle $ for the ground state
of Hamiltonian $\tilde{H}$ in Fig. \ref{boundary}. The Fig. \ref{boundary}%
(a,b) show the boundary lines as a function of detunings $\Delta _{1}$ and $%
\Delta _{2}$, and there exist an optimal detuning value which induces the MI
lobes with the most obvious separation. Any deviation from this optimal
detuning $\Delta _{1}$ ($\Delta _{2}$) will lead to symmetric (asymmetric)
convergence of the MI lobes whose polariton number greater than one. From
Fig. \ref{boundary}(c,d), we find that the phenomena of QPT disappear once
on-site repulsive interaction turns off through setting $\Omega =0$ or $g=0$%
.

Taking account of dissipation effects of polariton states, which resulting
from the decays of both cavity fields and NVCs, we simulate the
non-equilibrium dynamics by integrating the MME with the following form

\begin{equation}
\dot{\rho}=-i[H^{m},\rho ]+\frac{\kappa }{2}D[a]\rho +\sum\nolimits_{i=1,2}%
\frac{\Gamma _{i}}{2}D[\Gamma _{i}]\rho ,
\end{equation}%
where $D[A]\rho =2A\rho A^{+}-A^{+}A\rho -\rho A^{+}A$, $\kappa $ is the
nanocavity drcay rate, and $\Gamma _{1}$ ($\Gamma _{2}$) are the spontaneous
decay rates from excited state $|e\rangle $ to ground states $|g_{1}\rangle $
($|g_{2}\rangle $). The dissipative phase diagram is shown in Fig. \ref%
{mott-sf_diss}. It is found that the MI-lobe structure gradually disappear
and the area of MI phase expands as the dissipation strength increases,
i.e., new MI phase forms in the SF phase region existing in the
dissipationless case. The values of $\psi _{a}$ also decrease when the
dissipation effects are considered. In the MI phase, the dissipation has a
greater influence on the region with higher $\langle {N}\rangle $. For a
certain $\mu $, the value of $\psi _{a}$ near the phase boundary is bigger
than that away from the phase boundary. In contrast, the dissipation effect
slightly increase $\psi _{a}$ in the SF phase. Surprisingly, we observe that
there exists an oscillatory region in the SF phase region. This is due to
the emergence of multi-steady state resulting from the mean-field
approximation \cite{C3,PhysRevA.93.023821}, and this multi-stability will
disappear once the spatial quantum correlations are considered.

\textit{Detection of the QPT.-} We pay particular attention to using the
physical observable quantities to detect the QPT in the present system. Note
that the global signatures of the transition, such as the order parameter
and compressibility, are revealed in the photon number. Therefore we study
the critical behavior of mean intracavity photon number $\langle {n_{a}}%
\rangle $, photon fluctuation $\langle (\Delta {n_{a}})^{2}\rangle $ and
2nd-order equal-time correlation function $g^{(2)}(0)$, which allows one to
identify the optical signatures of the QPT, as shown in Fig. \ref{phofunc},
where we plot $\langle n_{a}\rangle $, $\langle (\Delta n_{a})^{2}\rangle $,
and $g^{(2)}(0)$ as a function of hopping rate $k$ for different $\mu $,
computed from steady state solutions of the MME. In Fig. \ref{phofunc}(a),
the photon number $\langle {n_{a}}\rangle $ is integer (non-integer) in the
MI (SF) phase in the dissipationless case. In contrast, $\langle {n_{a}}%
\rangle $ always decays to zero in MI phase, and quickly converges in SF
phase when $k$ increases in the dissipation case (Fig. \ref{phofunc}(d)).
The Fig. \ref{phofunc}(b,e) reveal that quantum fluctuations arising from
Heisenberg's uncertainty relation drive the transition from MI phase to SF
phase because the value of $\langle (\Delta n_{a})^{2}\rangle $ is zero
(nonzero) in MI (SF) phase. In the dissipationless case (Fig. \ref{phofunc}%
(b)), the maximal fluctuation occurs near the phase boundary and undergoes a
discontinuous change with a cusp-like character, then converges to a finite
value at large $k$. In the dissipation case (Fig. \ref{phofunc}(e)), the
fluctuation abruptly arises at the phase boundary and gradually converges at
a larger value compared with the dissipationless case.
\begin{figure}[tbp]
\begin{tabular}{cc}
\includegraphics[width=8.5cm]{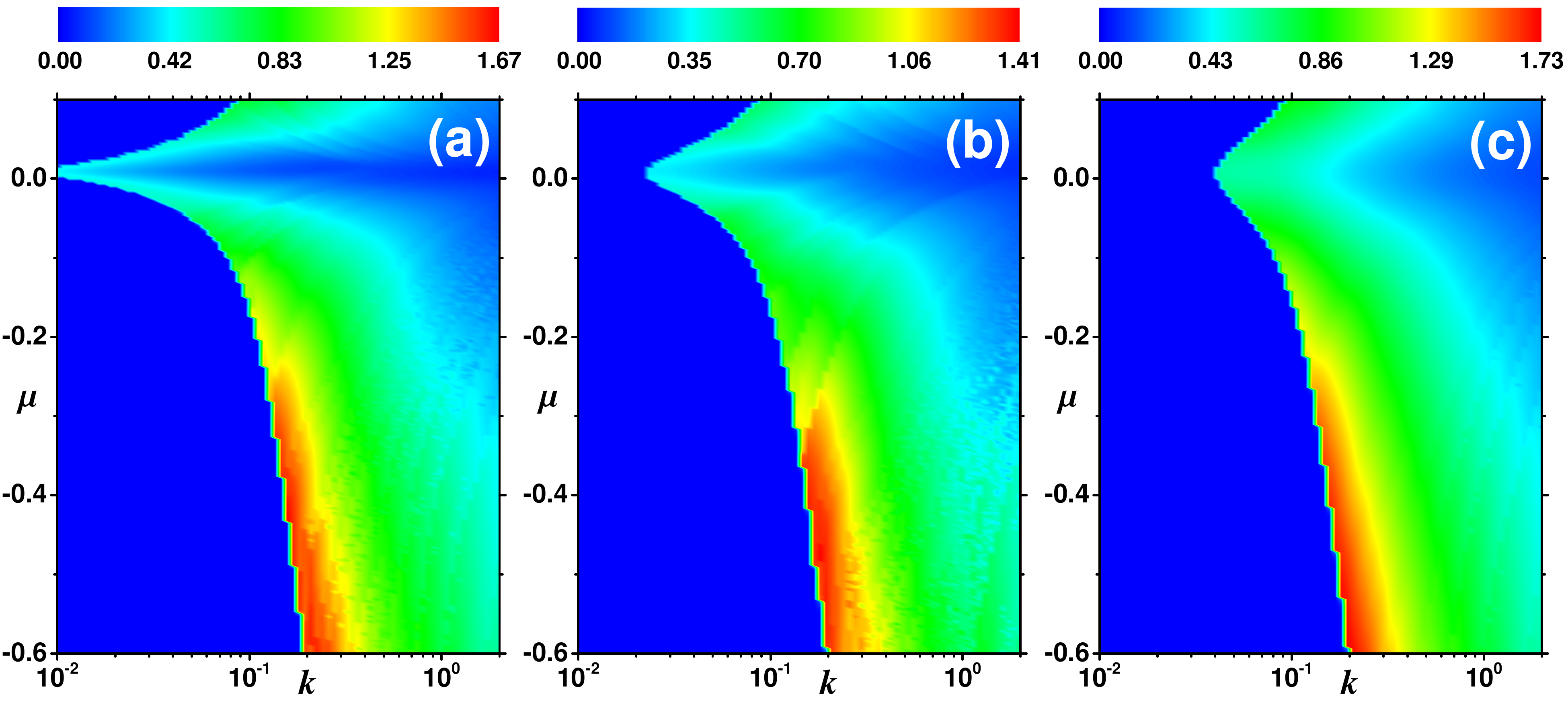} &
\end{tabular}%
\caption{Disspative quantum phase diagrams in the $\protect\mu \sim k$ plane
for different parameters.  $\Omega =5$, $\Delta _{1}=4$, $\Delta
_{2}=-2.5$; (a) $\Gamma _{1}=\Gamma _{2}=0.01$, $\protect\kappa =0.01$; (b) $%
\Gamma _{1}=\Gamma _{2}=0.05$, $\protect\kappa =0.01$; (c) $\Gamma _{1}=\Gamma
_{2}=0.05$, $\protect\kappa =0.05$.}
\label{mott-sf_diss}
\end{figure}

The density correlation function defined by $g^{(2)}(0)=\langle a^{\dag
}a^{\dag }aa\rangle /\langle a^{\dag }a\rangle ^{2}$ also exhibits distrinct
behaviors in different phases. In the dissipationless case (Fig. \ref%
{phofunc}(c)), we find that $g^{(2)}(0)<1$ in MI phase indicates photon
antibunching with sub-Poissonian statistics and photon blockade, and $%
g^{(2)}(0)$ gradually converges around 1 in SF phase with the growth of $k$.
The dissipation case (Fig. \ref{phofunc}(f)) exhibits a sudden transition of
$g^{(2)}(0)$\ from strong photon antibunching ($g^{(2)}(0)\ll 1$) to photon
bunching ($g^{(2)}(0)>1$) ith super-Poissonian statistics if we continuously
increase $k$. Note that $g^{(2)}(0)=1+[\left\langle (\Delta
n)^{2}\right\rangle -\left\langle n\right\rangle ]/\left\langle
n\right\rangle ^{2}<1$ due to zero variance $(\Delta n=0)$ and constant
photon number in MI phase, whereas $g^{(2)}(0)=1$ in SF phase because it
could be represented by a coherent state $\left\vert \alpha \right\rangle $
with non-integer polariton number and large fluctuation \cite{Gla}. In the
experiments one can infer the cavity field quadrature amplitudes,
intracavity photon number, and photon correlations by continuously
monitoring the output of PCC through available photon detectors \cite%
{det1,det2}. Additionally, the measurements of $g^{(2)}(0)$ could be
accessed by a modified heterodyne/homodyne or a Hanbury-Brown-Twiss setup
\cite{Walls}, and a recent experiment made a direct measurement of $%
g^{(2)}(0)$ on NVC by the characterization of fluorescent objects \cite{PRB}.%
\emph{\ }
\begin{figure}[tbp]
\begin{tabular}{cc}
\includegraphics[width=8cm]{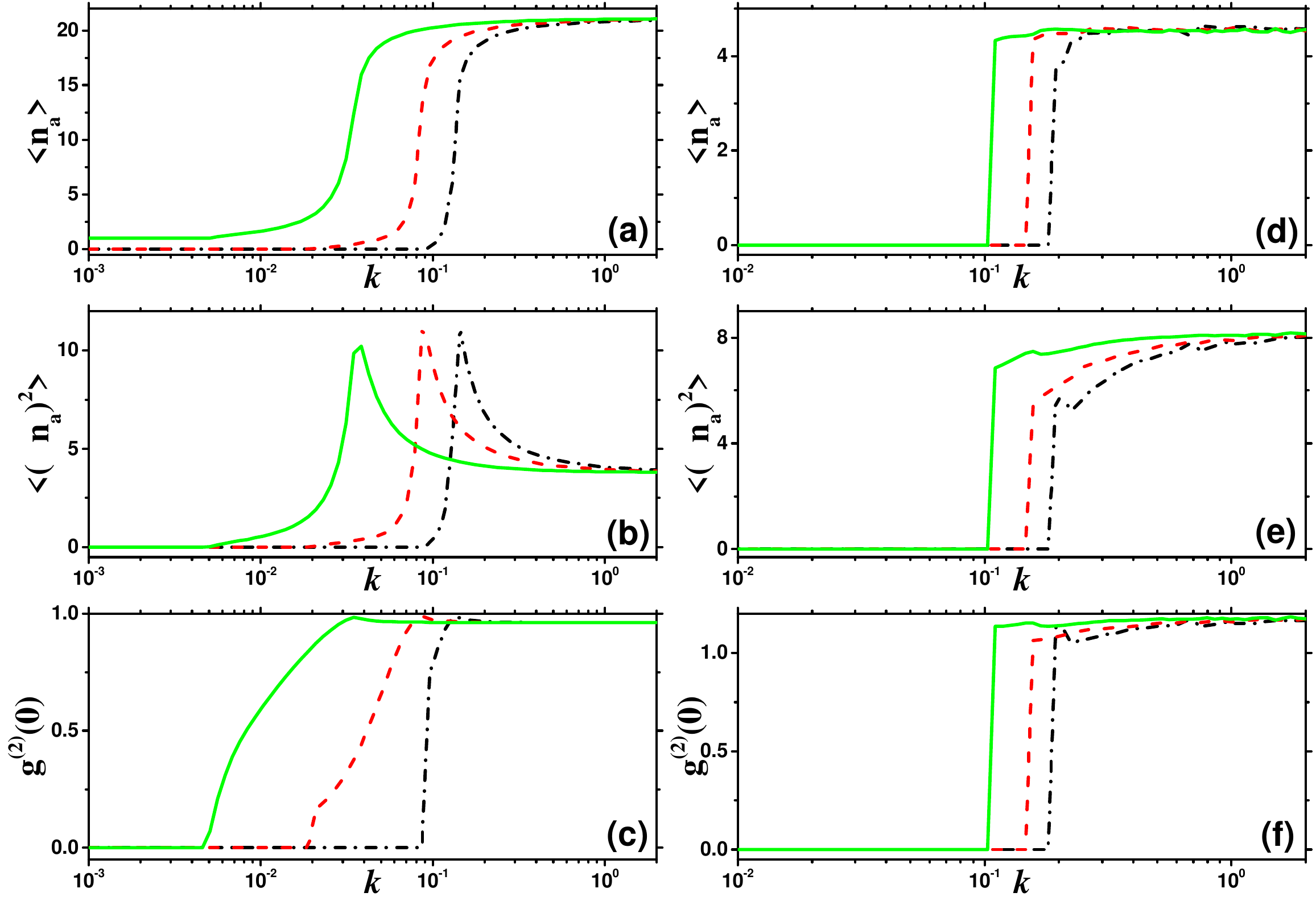} &  \\
&
\end{tabular}%
\caption{Mean intracavity photon number $\langle {n_{a}}\rangle $ (top),
photon fluctuation $\langle (\Delta {n_{a}})^{2}\rangle $ (middle) and
intensity correlation function $g^{(2)}(0)$ (bottom) as a function of
hopping rate $k$ for $\protect\mu =-0.6$ (black line)$,$ $-0.4$ (red line), $%
-0.2$ (green line). (a), (b), (c) are for dissipationless case with the same
parameters in Fig. (\protect\ref{mott-sf_origin}a). (d), (e), (f) are for
dissipation case with the same parameters in Fig. (\protect\ref{mott-sf_diss}%
a).}
\label{phofunc}
\end{figure}

The normalized emitted spectrum (NES) of system is another excellent optical
signature to detect the different phases.\textbf{\ }Next we turn to study
the two-time correlation function, whose spectral counterpart corresponds to
a concrete readily measurable quantity. Specifically, we show how to detect
the MI-SF phase transition for this dissipative-driven system, through
distinguishing NES in different phases.\emph{\ }It is convenient to
calculate the NES by combining Wiener-Khintchine theorem \cite{theo1,theo2}
with MME, and it can be written as
\begin{equation}
S_{\alpha }(\omega )=\frac{1}{\pi n_{\alpha }^{SS}}\lim_{t\rightarrow \infty
}\text{Re}\int_{0}^{\infty }G_{\alpha }^{(1)}(t,\tau ){e^{i\omega \tau }}%
d\tau ,
\end{equation}%
where the first-order time autocorrelator is $G_{\alpha }^{(1)}(t,\tau
)=\left\langle \alpha ^{\dagger }(t)\alpha (t+\tau )\right\rangle
-\left\vert \left\langle \alpha (t)\right\rangle \right\vert ^{2}$, and $%
n_{\alpha }^{SS}$ is the steady-state photon number with $\alpha =a,\sigma
_{1}^{-},\sigma _{2}^{-}$ \cite{PRL}.

\begin{figure}[tbp]
\begin{tabular}{cc}
\includegraphics[width=4cm]{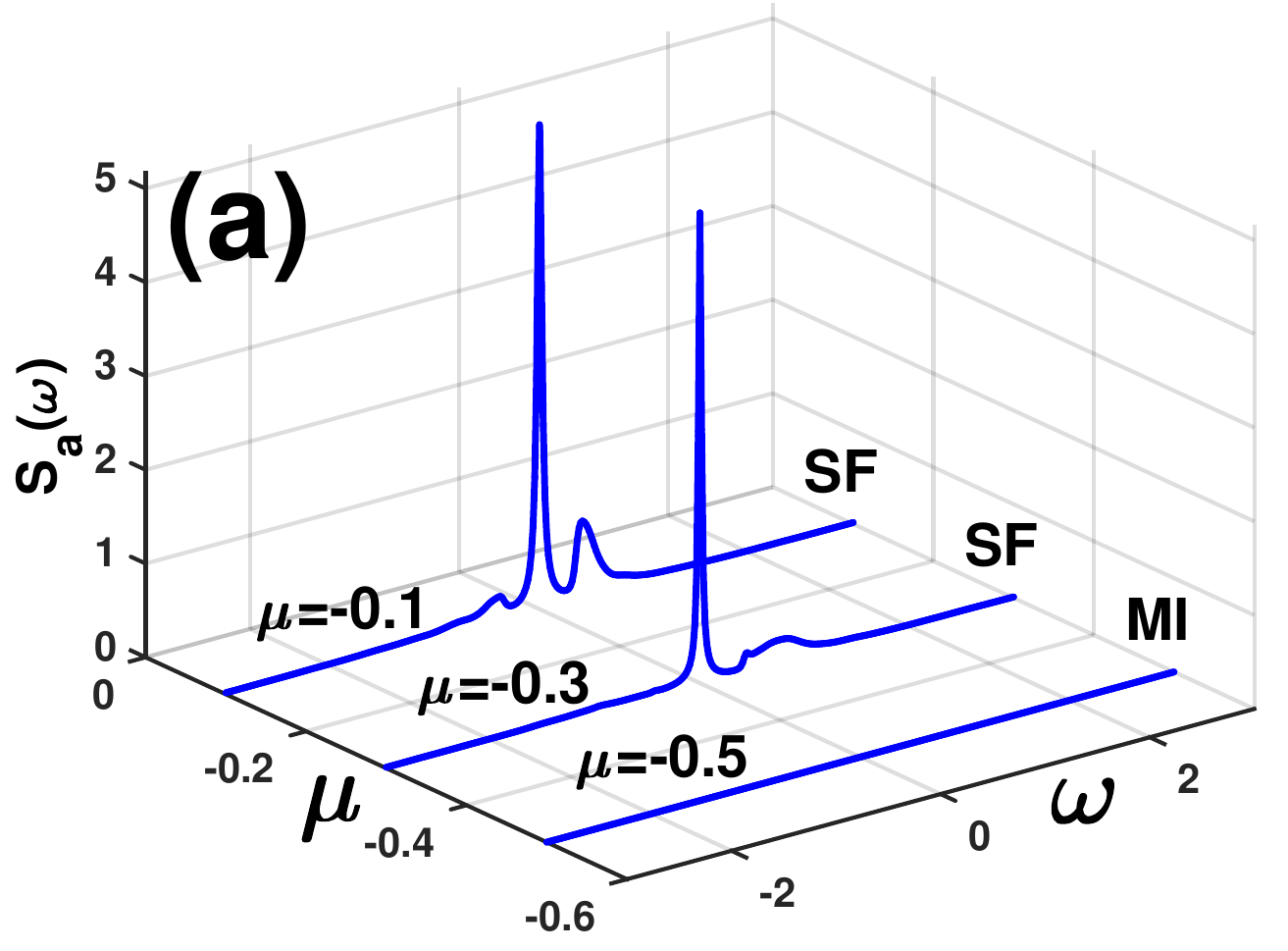} & %
\includegraphics[width=4cm]{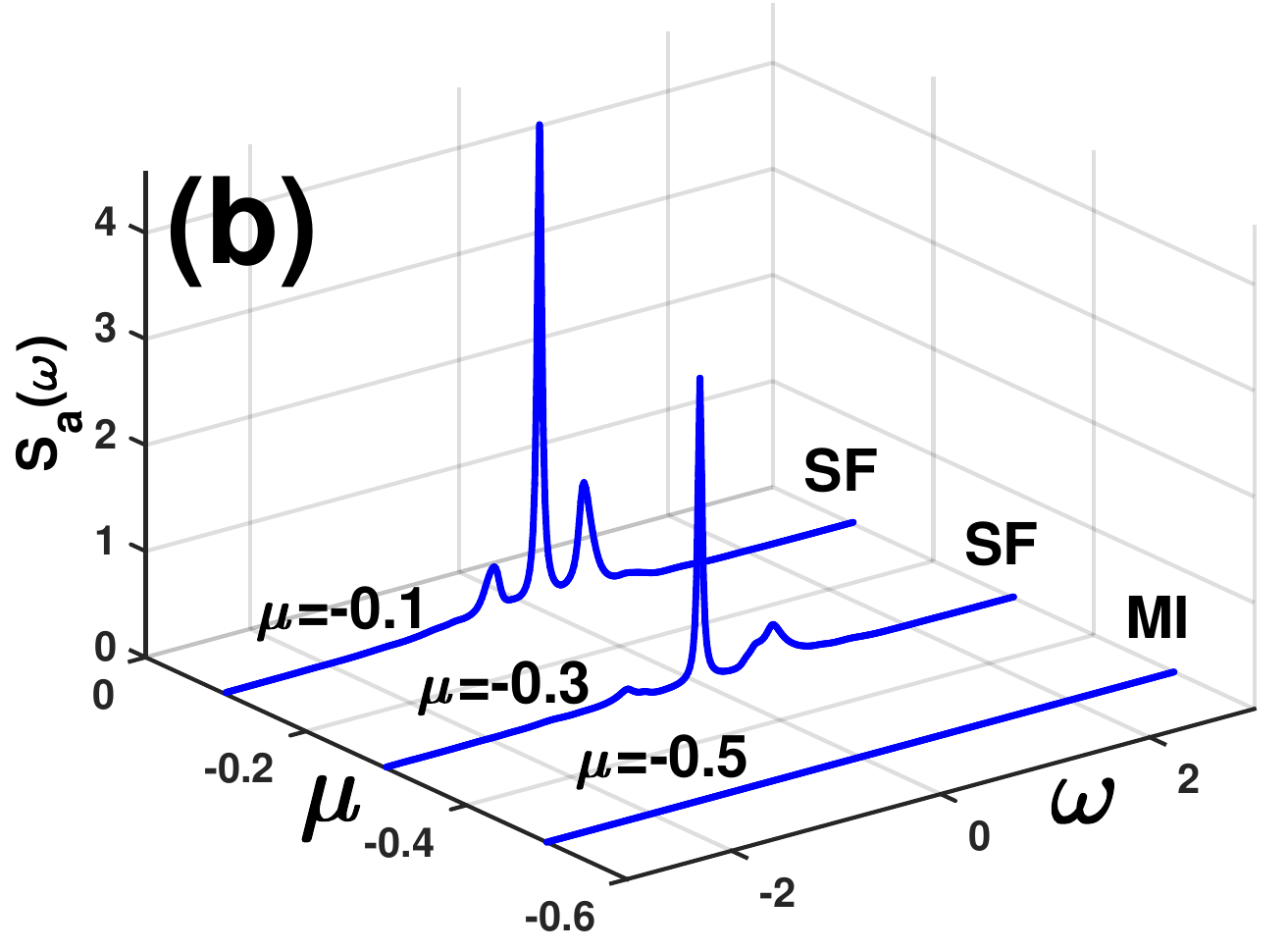} \\
\includegraphics[width=4cm]{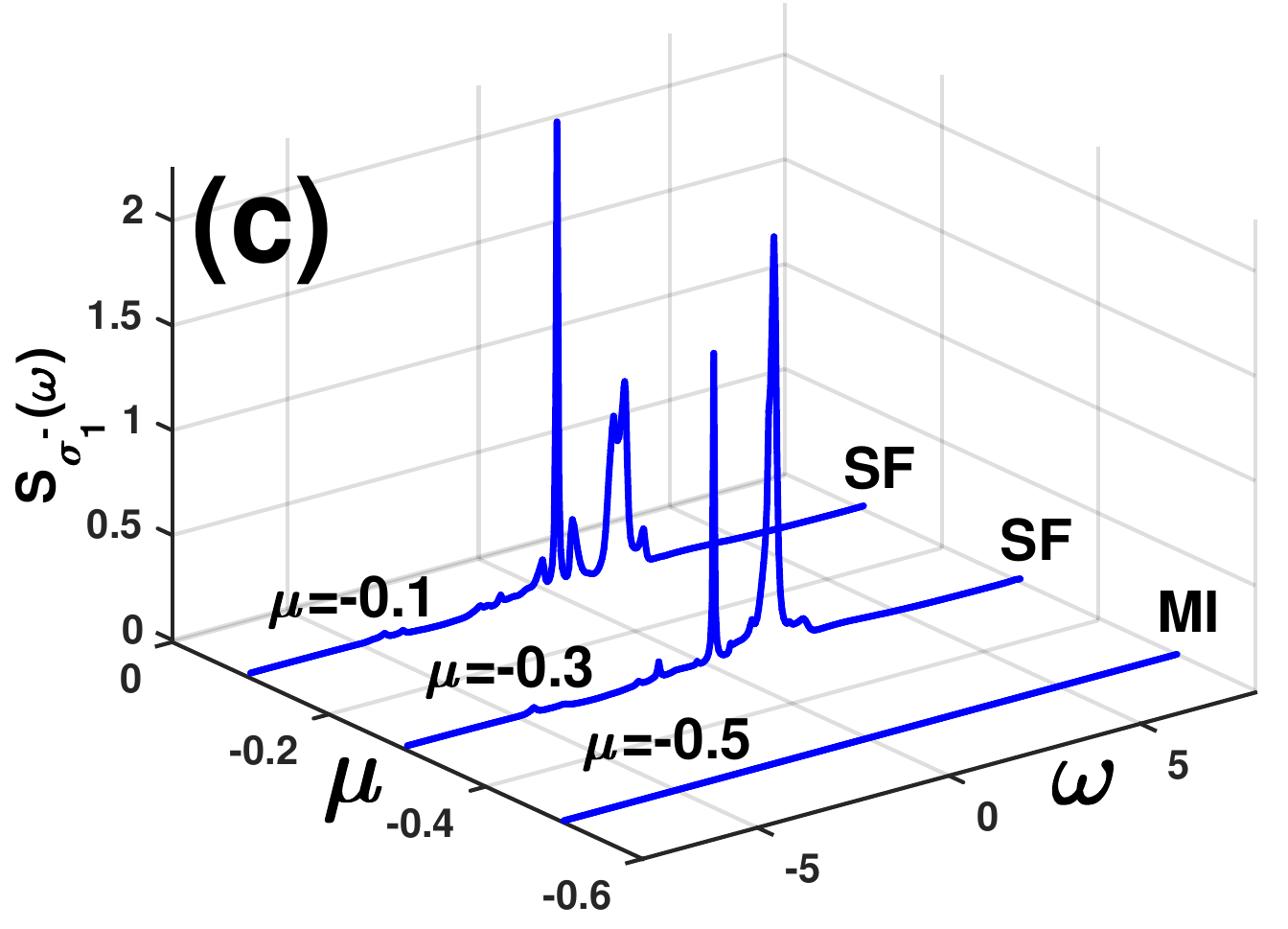} & %
\includegraphics[width=4cm]{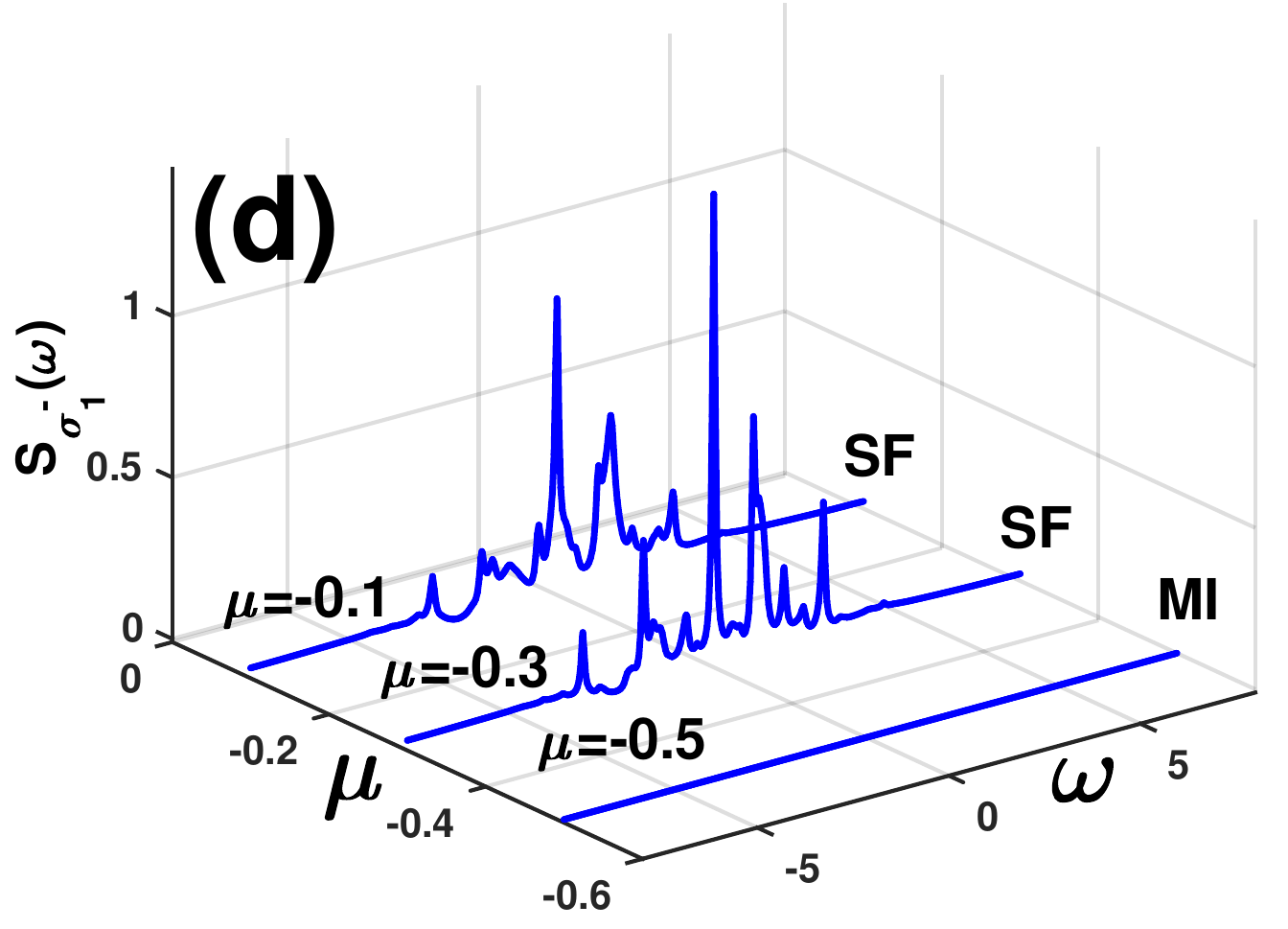} \\
\end{tabular}%
\caption{NES of nanocavity in the different phases. The parameters are (a,c) $\Delta_{1}=4$, $\Delta_{2}=-2.5$, $\Omega=0$, $\eta=0.05$, $k=0.13$, $\mu=-0.1,-0.3,-0.5$; (b,d) $\Delta_{1}=4$, $\Delta_{2}=-2.5$, $\Omega=5$, $\eta=0.05$, $k=0.13$, $\mu=-0.1,-0.3,-0.5$.}
\label{emitted_spectrum}
\end{figure}

Note that in the disspation case the steady photon number $n_{\alpha
}^{SS}=0 $ when the system is in MI phase (Fig. \ref{phofunc}(d)), there is
nothing could be observed in the NES. In contrast, the hopping term under
mean-field approximation in Eq. (\ref{mean_field}) is similar to a coherent
pumping when $\psi _{a}$ is nonzero in SF phase.\emph{\ }Fig. \ref%
{emitted_spectrum} show the NES of nanocavity in different phases, where the
lineshape of NES of nanocavity $S_{a}(\omega )$ is the standard "Mollow
triplet" ("Straight line") in SF (MI) phase (Fig. \ref{emitted_spectrum}(a,b)), and the intensity and location of
the sideband peaks can be changed by the chemical potential. Therfore, the
NES of nanocavity could also be a good indicator of dissipative QPT.

\textit{Summary}.- In conclusion, we propose a composite NVC-PCC system for
engineering a photonic QPT in a well-controllable way, where the effective
on-site repulsion can be tuned by changing the laser frequency and
intensity, while the cavity frequency and hopping strength could be adjusted
by the geometrical parameters of the defects. The NVCs remains\ individually
addressable, full control at the single-particle level, and site-resolved
measurement. The physical behind MI-SF phase transition is that quantum
fluctuations arising from Heisenberg's uncertainty relation drive the
transition. We also focus on using several experimentally observable
quantities exhibiting distinct optical signatures in different quantum
phases to detect the localization-delocalization transition. Our work opens
new perspectives in quantum simulation of condensed-matter and many-body
physics using such a 2D spin-cavity system.

This work was supported by the National Key Research and Development Program
of China under Grant No. 2017YFA0304503 and by the National Natural Science
Foundation of China under Grant No. 11574353. The IHPC A*STAR Team would
like to acknowledge the National Research Foundation Singapore (Grant No.
NRF2017NRF-NSFC002-015, NRF2016-NRF-ANR002, NRF-CRP 14-2014-04) and A*STAR
SERC (Grant No. A1685b0005).


%

\end{document}